\documentstyle{article}
\def\abstract#1{\vskip 7mm 
        \begin{center}{\large Abstract}\par \smallskip
                \begin{minipage}[c]{12cm}
                        \small #1
                \end{minipage}
        \end{center}
}
\def\title#1{\begin{center}{\Large\bf #1}\end{center}}
\def\author#1{\vskip 5mm \begin{center}{#1}\end{center}}
\def\address#1{\begin{center}{\it #1}\end{center}}

\newtheorem{theorem}     {Theorem}
\newtheorem{lemma}       {Lemma}

\newtheorem{corollary}   {Corollary}

\newcommand{\dalm}{\kern1pt\vbox{\hrule height 0.9pt
\hbox{\vrule width 0.9pt\hskip 2.5pt\vbox{\vskip 5.5pt}
\hskip 3pt\vrule width 0.3pt}\hrule height 0.3pt}\kern1pt}

\begin{document}
\begin{center}

\end{center}
\title{On initial conditions and global existence 
for accelerating cosmologies from string theory}

\author{Makoto Narita\footnote{Present address: 
Center for Relativity and Geometric Physics Studies, 
Department of Physics, 
National Central University, 
Jhongli 320, Taiwan 
Electronic address: narita@phy.ncu.edu.tw
}} 

\address{Max-Planck-Institut f\"ur Gravitationsphysik, \\
Albert-Einstein-Institut, 
Am M\"uhlenberg 1, 
D-14476 Golm, \\
Germany\\
E-mail: maknar@aei-potsdam.mpg.de
}

\abstract{
We construct a solution satisfying initial 
conditions for accelerating cosmologies 
from string/M-theory. 
Gowdy symmetric spacetimes with a positive potential are considered. 
Also, a global existence theorem for the spacetimes is shown. 
}

\section{Introduction}
\label{intro}
It is expected that the inflation paradigm would be explained within 
superstring/M-theory. The theory predicts that spacetime dimension is greater 
than four. 
Since observable spacetime dimension is four, it is thought that 
the extra dimensions would be compactified within Planck scale. 
Recently, it has been pointed out that 
it is possible to find cosmological 
solutions which exhibit a transient phase of 
accelerated expansion of the universe (like inflation) 
if the size of the 
compactified internal hyperbolic space 
depends on time and/or if they are S(pacelike)-brane 
solutions~\cite{EG,TP,WMNR}. 
In these models, exponential potential terms like $V_{0}e^{a\psi}$ appear, 
where $\psi$ denotes the compactification volume or effective dilaton field, 
$a$ is a coupling constant   
and $V_{0}$ is positive number. 
Explicitly, a typical action for the case is of the form 
\begin{eqnarray}
\label{typical-action}
S=\int d^{4}x\sqrt{-g}\left[-\;^{4}\!R
+\frac{1}{2}(\nabla\psi)^{2}
+V_{0}e^{a\psi}
\right].
\end{eqnarray}
Then, it is explained that 
if it would be supposed 
that, in the case of $a>0$, the field $\psi$ starts at a large negative value 
(i.e. the potential term can be neglected) with high kinetic energy 
($\partial_{t}\psi$ is positive and large enough)
\footnote{In the case of $a<0$, $\psi$ and 
$\partial_{t}\psi$ start at large positive and negative values, 
respectively.}
near cosmological initial 
singularities, 
then, the scalar field runs up the exponential potential, 
turn around and falls back. 
At the turning point, 
the potential term becomes dominant, i.e. the universe makes 
accelerated expansion. 
Thus, the universe starts out 
in a decelerated expansion phase (asymptotic past) 
and enters an accelerating phase (intermediate era), 
and after these, the expansion becomes deceleration again 
(asymptotic future).  
We call this scenario paradigm-A.

We would like to investigate this interesting paradigm from viewpoint of 
mathematical relativity and cosmology.  
It is important to study rigorously whether or not 
the paradigm-A is acceptable. 
In particular, it should be shown that 
the assumption of the initial conditions for $\psi$ is generic because, 
as indicated previously~\cite{EG}, the accelerated expansion of the universe 
is all the result of the initial conditions. 
That is, 
(Q1): Are there singular 
solutions satisfying 
initial conditions in paradigm-A to the Einstein-matter equations in generic? 

Furthermore, 
to be complete the scenario of paradigm-A, 
we should show global existence theorems, i.e., 
(Q2): Are there global solutions to the Einstein-matter equations with 
such exponential potentials? 
Unlike BKL~\cite{BKL} or cosmic no-hair conjectures~\cite{WR}, 
which are problems in only asymptotic 
(local) regions of spacetimes, 
the paradigm-A is a {\it global} (in time) problem as mentioned already. 
In addition, it is also important as the first step to prove the strong 
cosmic censorship.

For (Q1), 
to construct solutions satisfying the initial condition of paradigm-A, 
we will use the Fuchsian algorithm developed by Kichenassamy and 
Rendall~\cite{KR}. 
It is interpreted that the class of solutions we are looking for here is a 
subclass of asymptotically velocity-terms dominated (AVTD) 
singular solutions since potential terms are neglected 
near the singularities 
and, in addition, signature of the time derivative of the scalar field is 
restricted.  
By using the method, it has been shown that there are 
AVTD singularities in 
(non-)vacuum Gowdy, polarized $T^2$-, 
polarized $U(1)$-symmetric spacetimes and the 
Einstein-scalar-$p$-from system without symmetry assumptions
~\cite{AR,DHRW,IK,IM,NTM}. 
Also, systems with an exponential potential as given 
in (\ref{typical-action}) 
have been discussed formally in~\cite{DHRW,RA00}. 
Thus, our result is not only 
an answer for (Q1), but also it complements previous results.

For (Q2), we want to analyze Gowdy symmetric spacetimes. 
Future global existence theorems for spatially compact, locally homogeneous 
spacetimes~\cite{LH03,LH04,RA04} and hyperbolic symmetric spacetimes~\cite{TR} 
with a positive potential (or a positive cosmological constant)   
have been proved. These spacetimes do not include gravitational waves. 
Also, although 
global existence theorems for Gowdy (more generally, $T^2$-) symmetric 
spacetimes with or without matter have been 
shown~\cite{AH,ARW,BCIM,IW,MV,NM02,NM03,WM}, 
it has not been prove the theorems for the spacetimes 
with a positive potential. 
Therefore, spacetimes with dynamical 
degrees of freedom of gravity and with the positive potential 
should be considered as the next step.

As a model, we choose the bosonic action arising in 
low energy effective superstring 
(supergravity) theory since we have a similar action with 
(\ref{typical-action}) after the toroidal compactification of the 
extra dimensions.  
There are anti-symmetric two-form, $B_{\mu\nu}$, and three-form, 
$C_{\mu\nu\rho}$ fields in the action. 
It is known that, in general, 
$p$-form fields in $n$-dimensional spacetimes may violate 
the strong energy condition for $p\geq n-1$ and then, accelerated expansion 
of the universe would be expected~\cite{GG}. 
Here, 
we do not consider hyperbolic compactification of the extra higher dimensions, 
but the only fluxes of four-form field strengths are investigated because 
these have essentially 
the same effects to obtain the exponential potential terms 
as (\ref{typical-action})~\cite{EG,TP,WMNR}. 

Then, 
our purposes are to construct singular solutions satisfying conditions of 
paradigm-A and to show a global existence theorem for Gowdy symmetric 
spacetimes with stringy matter fields.

\subsection{Action}
The dimensionally reduced effective action in the Einstein frame is given by 
\begin{eqnarray}
\label{action-original}
S_{{\rm IIA}}=\int d^{4}x\sqrt{-g}\left[-\;^{4}\!R
+\frac{1}{2}(\nabla\phi)^{2}
+\frac{1}{2\cdot3!}e^{-2\lambda\phi}H^{2}
+\frac{1}{2\cdot4!}e^{-2\lambda\phi}F^{2}
\right],
\end{eqnarray}
where $g$ is the determinant of the metric $g_{\mu\nu}$ 
on a four-dimensional spacetime manifold $M$, 
$\;^{4}\!R$ is the Ricci scalar of $g_{\mu\nu}$, $\phi$ is the dilaton field, 
$H=dB$ is 
the three-form field strength, $F=dC$ is the 
four-form field strength and $\lambda$ is a coupling constant. 
If $\lambda=1$, we have the action for the type IIA supergravity 
in the absence of vector fields and the Chern-Simons term~\cite{LWC}. 

In four dimensions, there is a duality between the three-form field strength 
and a one-form, which is interpreted as the gradient of a scalar field. 
Then, we may define the pseudo-scalar axion field $\sigma$ as follows: 
\begin{eqnarray}
H^{\mu\nu\rho}=
\epsilon^{\mu\nu\rho\kappa}e^{2\lambda\phi}\nabla_{\kappa}\sigma.  
\end{eqnarray}
Also, the field equation 
\begin{eqnarray}
\nabla_{\mu}\left(e^{-2\lambda\phi}F^{\mu\nu\rho\kappa}\right)=0, 
\end{eqnarray}
and the Bianchi identity 
\begin{eqnarray}
\partial_{[\alpha}F_{\mu\nu\rho\kappa]}=0, 
\end{eqnarray}
for the four-form field strength 
can be solved by
\begin{eqnarray}
F^{\mu\nu\rho\kappa}=Q\epsilon^{\mu\nu\rho\kappa}e^{2\lambda\phi}, 
\end{eqnarray}
where $Q$ is an arbitrary constant. 
Thus, 
after taking the dual transformation and solving the field equations for $F$, 
we have a reduced effective action for the IIA system of the form 
\begin{eqnarray}
\label{action}
S_{{\rm IIA*}}=\int d^{4}x\sqrt{-g}\left[-\;^{4}\!R
+\frac{1}{2}\left\{(\nabla\phi)^{2}
+e^{2\lambda\phi}(\nabla\sigma)^{2}
+Q^{2}e^{2\lambda\phi}
\right\}\right].
\end{eqnarray}
Hereafter, we assume $Q\neq0$. Thus, we have the action which is the same 
from with (\ref{typical-action}).
\subsection{Field equations for Gowdy symmetric spacetimes}
The Gowdy symmetric spacetimes admit a $T^2$ isometry group 
with spacelike orbits and the twists associated to the group vanish~\cite{GR}. 
The topology of spatial section can be accepted $S^3$, $S^2\times S^1$, 
$T^3$ or the lens space~\cite{CP}. 
In this paper, we assume $T^3$ spacelike topology. 

Now, we will choose a coordinate, which is the {\it areal time} one. 
This means that time $t$ is proportional to the geometric area of the 
orbits of the isometry group. Explicitly, 
\begin{equation} 
\label{areal-metric}
ds=-e^{2(\eta-U)}\alpha dt^{2}+e^{2(\eta-U)}
d\theta^{2}+e^{2U}(dx+Ady)^{2}+e^{-2U}t^{2}dy^{2}\label{areal}, 
\end{equation}
where $\partial/\partial x$ and $\partial/\partial y$ are Killing vector 
fields generating the $T^2$ group action, and $\eta$, $\alpha$, $U$ and $A$ 
are functions of $t\in(0,\infty)$ and $\theta\in S^1$. It is also assumed that 
functions describing behavior of matter fields are ones of $t$ and $\theta$. 

Let us show the field equations obtained by varying the action (\ref{action}) 
in the 
areal coordinate~(\ref{areal-metric}). 

\noindent
{\it Constraint equations}
\begin{eqnarray}
\label{ce-eta}
\frac{\dot{\eta}}{t}=\dot{U}^{2}&+&\alpha U'^{2}+
\frac{e^{4U}}{4t^2}(\dot{A}^2+\alpha A'^2) \nonumber \\
&+&
\frac{1}{4}\left[\dot{\phi}^{2}+\alpha\phi'^{2}
+e^{2\lambda\phi}(\dot{\sigma}^{2}+\alpha\sigma'^{2})
+\alpha Q^{2}e^{2\lambda\phi+2(\eta-U)}\right],
\end{eqnarray}
\begin{eqnarray}
\label{ce-eta'}
\frac{\eta'}{t}=2\dot{U}U'+
\frac{e^{4U}}{2t^2}\dot{A}A'-\frac{\alpha'}{2t\alpha}
+\frac{1}{2}(\dot{\phi}\phi'+e^{2\lambda\phi}\dot{\sigma}\sigma'),
\end{eqnarray}
\begin{eqnarray}
\label{ce-alpha}
\dot{\alpha}=-t\alpha^{2}Q^{2}e^{2\lambda\phi+2(\eta-U)}.
\end{eqnarray}
{\it Evolution equations}
\begin{eqnarray}
\label{ee-eta} 
\ddot{\eta}-\alpha\eta''=&&\frac{\eta'\alpha'}{2}+
\frac{\dot{\eta}\dot{\alpha}}{2\alpha}-\frac{\alpha'^{2}}{4\alpha}
+\frac{\alpha''}{2}-\dot{U}^{2}+\alpha
U'^{2}+\frac{e^{4U}}{4t^2}(\dot{A}^2-\alpha A'^2) \nonumber \\
&+&\frac{1}{4}\left[-\dot{\phi}^{2}+\alpha\phi'^{2}
+e^{2\lambda\phi}(-\dot{\sigma}^{2}+\alpha\sigma'^{2})
+\alpha Q^{2}e^{2\lambda\phi+2(\eta-U)}\right],
\end{eqnarray}
\begin{eqnarray}
\label{ee-u}
\ddot{U}-\alpha U''=-\frac{\dot{U}}{t}+\frac{\dot{\alpha}\dot{U}}{2\alpha}+
\frac{\alpha'U'}{2}
+
\frac{e^{4U}}{2t^2}(\dot{A}^2-\alpha A'^2)
+\frac{1}{4}\alpha Q^{2}e^{2\lambda\phi+2(\eta-U)},
\end{eqnarray}
\begin{eqnarray}
\label{ee-a}
\ddot{A}-\alpha A''=\frac{\dot{A}}{t}+\frac{\dot{\alpha}\dot{A}}{2\alpha}+
\frac{\alpha'A'}{2}-4(\dot{A}\dot{U}
-\alpha A'U'),
\end{eqnarray} 
\begin{eqnarray}
\label{ee-phi}
\ddot{\phi}-\alpha \phi''=-\frac{\dot{\phi}}{t}
+\frac{\dot{\alpha}\dot{\phi}}{2\alpha}+\frac{\alpha'\phi'}{2}
+\lambda e^{2\lambda\phi}(\dot{\sigma}^2-\alpha\sigma'^2)
-\lambda\alpha Q^{2}e^{2\lambda\phi+2(\eta-U)},
\end{eqnarray} 
\begin{eqnarray}
\label{ee-sigma}
\ddot{\sigma}-\alpha \sigma''=-\frac{\dot{\sigma}}{t}
+\frac{\dot{\alpha}\dot{\sigma}}{2\alpha}+\frac{\alpha'\sigma'}{2}
-2\lambda(\dot{\phi}\dot{\sigma}-\alpha\phi'\sigma').
\end{eqnarray} 
Hereafter, dot and prime denote derivative with respect to $t$ and $\theta$, 
respectively. We will call this system of partial differential equations 
(PDEs) {\it Gowdy symmetric IIA system}. 
Note that these equations are not independent because the 
wave equation (\ref{ee-eta}) for $\eta$ can be derived from other equations. 
Indeed, there are only two dynamical degree of freedom (i.e. $U$ and $A$) 
in the Gowdy symmetric spacetimes. 
\section{Initial singularities}
\label{initial-singularity}
Consider the problem (Q1). 
To begin with a brief review of the Fuchsian algorithm, which is a method to 
construct exact singular solutions 
to a PDE system near a singularity ($t=0$). 
The algorithm is based on the following idea: 
near the singularity, decompose the singular formal solutions into 
a singular part, which depends on a number of arbitrary functions, 
and a regular part $u$. If the system can be written as 
a Fuchsian system of the form 
\begin{eqnarray}
\label{fuchsian-eq}
\left[D+{\cal N}(x)\right]u=tf(t,x,u,\partial_{x}u), 
\end{eqnarray} 
where $D:=t\partial_{t}$ and $f$ is a vector-valued regular function, 
then the following theorem can be apply:
\begin{theorem}{\rm \cite{KR}}
\label{kr-theorem}
Assume that ${\cal N}$ is an analytic matrix near $x=x_{0}$ such that 
there is a constant $C$ with $\Vert\Lambda^{{\cal N}}\Vert\leq C$ 
for $0<\Lambda<1$. In addition, suppose that $f$ is a locally Lipschitz 
function of $u$ and $\partial_{x}u$ which preserves analyticity in $x$ 
and continuity in $t$. Then, the Fuchsian system (\ref{fuchsian-eq}) 
has a unique solution in 
a neighborhood of $x=x_{0}$ and $t=0$ 
which is analytic in $x$ and continuous in 
$t$ and tend to zero as $t\rightarrow 0$. 
\end{theorem}
Thus, the regular part goes to zero and the singular part 
of the formal solution becomes an exact solution to the original PDE system 
near the singularity. 

Unlike the vacuum Gowdy case, the evolution 
equations~(\ref{ee-u})-(\ref{ee-sigma}) do not decouple 
from the constraint equations~(\ref{ce-eta})-(\ref{ce-alpha}), 
since they contain the function $\alpha$. 
Therefore, according to~\cite{IK}, we take equations 
(\ref{ce-eta}), (\ref{ce-alpha}), 
(\ref{ee-u})-(\ref{ee-sigma}) as effective evolution ones 
and (\ref{ce-eta'}) as 
the only effective constraint equation. 
This is not a standard setup for the initial-value problem for the 
Einstein-matter equations (see example~\cite{TM}). 
Therefore, it is not clear whether 
the initial-value problem for our case away from the singularity at $t=0$ 
has a unique solution or not, unless it is shown that 
the constraint (\ref{ce-eta'}) propagates. 

Let us show the local existence and uniqueness of our initial-value problem. 
We can obtain the following first-order system for 
$\vec{z}$ from the PDE system (\ref{ce-eta}), 
(\ref{ce-alpha}), 
(\ref{ee-u})-(\ref{ee-sigma}): 
\begin{eqnarray}
\partial_{t}\vec{z}=f(t,\theta,\vec{z},
\partial_{\theta}\vec{z}), 
\end{eqnarray} 
where $\vec{z}:=(U,\dot{U},U',A,\dot{A},A',\phi,\dot{\phi},\phi',
\sigma,\dot{\sigma},\sigma',\alpha,\eta)$. 
This means that the PDE system is of Cauchy-Kowalewskaya type. 
Thus, ignoring the constraint equation (\ref{ce-eta'}), we have a unique 
solution to the effective evolution equations by prescribing the analytic 
initial data for $t=t_{0}>0$ if all functions are analytic.  

Now,  
to assure the local existence and uniqueness of the initial-value problem, 
we must show that the constraint (\ref{ce-eta'}) propagates.  
Let us set 
\begin{eqnarray}
\label{N}
N:=\eta'-2DUU'-\frac{e^{4U}}{2t^2}DAA'-\frac{1}{2}D\phi\phi'
-\frac{e^{2\lambda\phi}}{2}D\sigma\sigma'+\frac{\alpha'}{2\alpha},
\end{eqnarray} 
Computing 
\begin{eqnarray}
0&=&D\eta'-(D\eta)'\nonumber \\ 
&=&DN+
D\left(2DUU'+\frac{e^{4U}}{2t^2}DAA'+\frac{1}{2}D\phi\phi'
\frac{e^{2\lambda\phi}}{2}D\sigma\sigma'-\frac{\alpha'}{2\alpha}\right)
-(D\eta)',
\end{eqnarray} 
we have a linear, homogeneous 
ordinary differential equation (ODE) for $N$ of the form 
\begin{eqnarray}
DN-\frac{D\alpha}{2\alpha}N=0.
\end{eqnarray} 
Thus, the uniqueness theorem for ODEs guarantees that $N$ is identically zero 
for any time $t$ if we set initial data for $t=t_{0}$ such that $N(t_{0})=0$. 
Thus, the local existence and uniqueness 
of the initial-value problem for our case 
has been shown in the analytic case.  
In appendix, we shall consider the smooth version of the initial-value problem 
for our non-standard setup of the Gowdy symmetric IIA system. 
\subsection{Application of the Fuchsian algorithm}
Let us construct AVTD singular solutions to the Gowdy symmetric IIA system. 
First, we will consider the case that a solution has a maximum number 
of free functions. In this sense, the solution (given in theorem~\ref{avtd1}) 
is generic. 

Neglecting spatial derivative and potential terms in the effective evolution 
equations, we have velocity-terms dominated (VTD) equations as follow:
\begin{eqnarray}
D\eta=(DU)^{2}+\frac{e^{4U}}{4t^2}(DA)^2+
\frac{1}{4}(D\phi)^{2}+\frac{e^{2\lambda\phi}}{4}(D\sigma)^2,
\end{eqnarray} 
\begin{eqnarray}
D\alpha=0,
\end{eqnarray} 
\begin{eqnarray}
D^2U=\frac{1}{2\alpha}DUD\alpha+\frac{e^{4U}}{4t^2}(DA)^2,
\end{eqnarray} 
\begin{eqnarray}
D^2A=2DA+\frac{1}{2\alpha}DAD\alpha-4DUDA,
\end{eqnarray} 
\begin{eqnarray}
D^2\phi=\frac{1}{2\alpha}D\phi D\alpha+\lambda e^{2\lambda\phi}(D\sigma)^2,
\end{eqnarray} 
\begin{eqnarray}
D^2\sigma=\frac{1}{2\alpha}D\sigma D\alpha-2\lambda D\phi D\sigma.
\end{eqnarray} 
Solving this system of VTD equations, 
we have a VTD solution. Then, the following formal solution is obtained:  
\begin{eqnarray}
\label{vtd-sol-eta}
\eta=\left(k(\theta)^2+\frac{\kappa(\theta)^2}{4}\right)\ln t+\eta_{0}(\theta)
+t^{\epsilon}\mu(t,\theta),
\end{eqnarray} 
\begin{eqnarray}
\label{vtd-sol-alpha}
\alpha=\alpha_{0}(\theta)+t^{\epsilon}\beta(t,\theta),
\end{eqnarray} 
\begin{eqnarray}
\label{vtd-sol-u}
U=k(\theta)\ln t +U_{0}(\theta)+t^{\epsilon}V(t,\theta),
\end{eqnarray} 
\begin{eqnarray}
\label{vtd-sol-a}
A=h(\theta)+t^{2-4k}\left(A_{0}(\theta)+B(t,\theta)\right),
\end{eqnarray} 
\begin{eqnarray}
\label{vtd-sol-phi}
\phi=\kappa(\theta)\ln t +\phi_{0}(\theta)+t^{\epsilon}\Phi(t,\theta),
\end{eqnarray} 
\begin{eqnarray}
\label{vtd-sol-sigma1}
\sigma=\omega(\theta)+t^{-2\lambda\kappa}\left(\sigma_{0}(\theta)
+\Sigma(t,\theta)\right),
\end{eqnarray} 
where 
\begin{eqnarray}
\label{velocity-condition}
\epsilon>0,\hspace{.5cm}0<k(\theta)<\frac{1}{2},\hspace{.5cm} 
\alpha_{0}>0
\end{eqnarray} 
and 
\begin{eqnarray}
\label{velocity-condition'}
-1<\lambda\kappa(\theta)<0. 
\end{eqnarray} 
Note that $\mu$, $\beta$, $V$, $B$, $\Phi$ and $\Sigma$ are regular parts and 
others are singular parts ($=$VTD solutions).

Inserting this formal solution into the Einstein-matter equations, we obtain 
the following Fuchsian system:
\begin{eqnarray}
\label{fuchsian-eq-gowdy}
(D+{\cal N})\vec{u}=t^{\delta}f(t,\theta,\vec{u},\partial_{\theta}\vec{u}),
\end{eqnarray} 
where $\vec{u}:=u_{i}=(V,DV,t^{\epsilon}V',B,DB,t^{\epsilon}B',
\Phi,D\Phi,t^{\epsilon}\Phi',\Sigma,D\Sigma,t^{\epsilon}\Sigma',\beta,\mu)$, 
$i=1,\cdot\cdot\cdot, 14$, 
$f$ is a vector-valued regular function and 
\begin{eqnarray}
{\cal N}=
\left(\begin{array}{cccccccccccccc}
0 & -1 & 0 & 0  & 0 & 0 & 0 & 0 & 0 & 0 & 0 & 0 & 0 & 0\\
\epsilon^2 & 2\epsilon & 0 & 0 & 0 & 0 & 0 & 0 & 0 & 0 & 0 & 0 & 0 & 0\\
0 & 0 & 0 & 0 & 0 & 0 & 0 & 0 & 0 & 0 & 0 & 0 & 0 & 0\\
0 & 0 & 0 & 0 & -1 & 0 & 0 & 0 & 0 & 0 & 0 & 0 & 0 & 0\\
0 & 0 & 0 & 0 & 2-4k & 0 & 0 & 0 & 0 & 0 & 0 & 0 & 0 & 0\\
0 & 0 & 0 & 0 & 0 & 0 & 0 & 0 & 0 & 0 & 0 & 0 & 0 & 0\\
0 & 0 & 0 & 0 & 0 & 0 & 0 & -1 & 0 & 0 & 0 & 0 & 0 & 0\\
0 & 0 & 0 & 0 & 0 & 0 & \epsilon^2 & 2\epsilon & 0 & 0 & 0 & 0 & 0 & 0\\
0 & 0 & 0 & 0 & 0 & 0 & 0 & 0 & 0 & 0 & 0 & 0 & 0 & 0\\
0 & 0 & 0 & 0 & 0 & 0 & 0 & 0 & 0 & 0 & -1 & 0 & 0 & 0\\
0 & 0 & 0 & 0 & 0 & 0 & 0 & 0 & 0 & 0 & -2\lambda\kappa & 0 & 0 & 0\\
0 & 0 & 0 & 0 & 0 & 0 & 0 & 0 & 0 & 0 & 0 & 0 & 0 & 0\\
0 & 0 & 0 & 0 & 0 & 0 & 0 & 0 & 0 & 0 & 0 & 0 & \epsilon & 0\\
-2k\epsilon & -2k & 0 & 0 & 0 & 0 & -\frac{\kappa\epsilon}{2} 
& -\frac{\kappa}{2} & 0 & 0 & 0 & 0 & 0 & \epsilon \\
\end{array}\right).
\end{eqnarray} 
Note that $\delta>0$ 
if the condition~(\ref{velocity-condition}), (\ref{velocity-condition'})
and 
\begin{eqnarray}
\label{velocity-condition2}
K:=(k-\frac{1}{2})^2+\frac{\kappa^2}{4}+\lambda\kappa+\frac{3}{4}>0
\end{eqnarray} 
holds.

To apply theorem~\ref{kr-theorem} to 
our Fuchsian system~(\ref{fuchsian-eq-gowdy}), 
we must to verify that the boundedness condition for the matrix ${\cal N}$ 
holds. 
To do this, we have $P^{-1}{\cal N}P={\cal N}_{0}$, where
\begin{eqnarray}
{\cal N}_{0}=
\left(\begin{array}{cccccccccccccc}
\epsilon & 1 & 0 & 0  & 0 & 0 & 0 & 0 & 0 & 0  & 0 & 0 & 0 & 0\\
0 & \epsilon & 0 & 0 & 0 & 0 & 0 & 0 & 0 & 0  & 0 & 0 & 0 & 0\\
0 & 0 & 0 & 0 & 0 & 0 & 0 & 0 & 0 & 0  & 0 & 0 & 0 & 0\\
0 & 0 & 0 & 0 & 0 & 0 & 0 & 0 & 0 & 0  & 0 & 0 & 0 & 0\\
0 & 0 & 0 & 0 & 2-4k & 0 & 0 & 0 & 0 & 0  & 0 & 0 & 0 & 0\\
0 & 0 & 0 & 0 & 0 & 0 & 0 & 0 & 0 & 0  & 0 & 0 & 0 & 0\\
0 & 0 & 0 & 0 & 0 & 0 & \epsilon & 1 & 0 & 0  & 0 & 0 & 0 & 0 \\
0 & 0 & 0 & 0 & 0 & 0 & 0 & \epsilon & 0 & 0 & 0 & 0 & 0 & 0\\
0 & 0 & 0 & 0 & 0 & 0 & 0 & 0 & 0 & 0 & 0 & 0 & 0 & 0\\
0 & 0 & 0 & 0 & 0 & 0 & 0 & 0 & 0 & 0 & 0 & 0 & 0 & 0\\
0 & 0 & 0 & 0 & 0 & 0 & 0 & 0 & 0 & 0 & -2\lambda\kappa & 0 & 0 & 0\\
0 & 0 & 0 & 0 & 0 & 0 & 0 & 0 & 0 & 0 & 0 & 0 & 0 & 0\\
0 & 0 & 0 & 0 & 0 & 0 & 0 & 0 & 0 & 0 & 0 & 0 & \epsilon & 0\\
0 & 2k & 0 & 0 & 0 & 0 & 0 & \frac{\kappa}{2} & 0 & 0 & 0 & 0 & 0 & \epsilon \\
\end{array}\right),
\end{eqnarray} 
and 
\begin{eqnarray}
P=
\left(\begin{array}{cccccccccccccc}
1 & 0 & 0 & 0  & 0 & 0 & 0 & 0 & 0 & 0  & 0 & 0 & 0 & 0\\
-\epsilon & -1 & 0 & 0 & 0 & 0 & 0 & 0 & 0 & 0  & 0 & 0 & 0 & 0\\
0 & 0 & 1 & 0 & 0 & 0 & 0 & 0 & 0 & 0  & 0 & 0 & 0 & 0\\
0 & 0 & 0 & 1 & 0 & 0 & 0 & 0 & 0 & 0  & 0 & 0 & 0 & 0\\
0 & 0 & 0 & 0 & 1 & 0 & 0 & 0 & 0 & 0  & 0 & 0 & 0 & 0\\
0 & 0 & 0 & 0 & 0 & 1 & 0 & 0 & 0 & 0  & 0 & 0 & 0 & 0\\
0 & 0 & 0 & 0 & 0 & 0 & 1 & 0 & 0 & 0  & 0 & 0 & 0 & 0 \\
0 & 0 & 0 & 0 & 0 & 0 & -\epsilon & -1 & 0 & 0 & 0 & 0 & 0 & 0 \\
0 & 0 & 0 & 0 & 0 & 0 & 0 & 0 & 1 & 0 & 0 & 0 & 0 & 0 \\
0 & 0 & 0 & 0 & 0 & 0 & 0 & 0 & 0 & 1 & 0 & 0 & 0 & 0 \\
0 & 0 & 0 & 0 & 0 & 0 & 0 & 0 & 0 & 0 & 1 & 0 & 0 & 0 \\
0 & 0 & 0 & 0 & 0 & 0 & 0 & 0 & 0 & 0 & 0 & 1 & 0 & 0 \\
0 & 0 & 0 & 0 & 0 & 0 & 0 & 0 & 0 & 0 & 0 & 0 & 1 & 0\\
0 & 0 & 0 & 0 & 0 & 0 & 0 & 0 & 0 & 0 & 0 & 0 & 0 & 1 \\
\end{array}\right).
\end{eqnarray} 
Then, 
\begin{eqnarray}
\Lambda^{{\cal N}_{0}}=
\left(\begin{array}{cccccccccccccc}
\Lambda^{\epsilon} & \Lambda^{\epsilon}\ln\Lambda & 0 
& 0  & 0 & 0 & 0 & 0 & 0 & 0  & 0 & 0 & 0 & 0\\
0 & \Lambda^{\epsilon} & 0 & 0 & 0 & 0 & 0 & 0 & 0 & 0  & 0 & 0 & 0 & 0\\
0 & 0 & 1 & 0 & 0 & 0 & 0 & 0 & 0 & 0  & 0 & 0 & 0 & 0\\
0 & 0 & 0 & 1 & 0 & 0 & 0 & 0 & 0 & 0  & 0 & 0 & 0 & 0\\
0 & 0 & 0 & 0 & \Lambda^{2-4k} & 0 & 0 & 0 & 0 & 0  & 0 & 0 & 0 & 0\\
0 & 0 & 0 & 0 & 0 & 1 & 0 & 0 & 0 & 0  & 0 & 0 & 0 & 0\\
0 & 0 & 0 & 0 & 0 & 0 & \Lambda^{\epsilon} & \Lambda^{\epsilon}\ln\Lambda & 0 
& 0  & 0 & 0 & 0 & 0 \\
0 & 0 & 0 & 0 & 0 & 0 & 0 & \Lambda^{\epsilon} & 0 & 0 & 0 & 0 & 0 & 0\\
0 & 0 & 0 & 0 & 0 & 0 & 0 & 0 & 1 & 0 & 0 & 0 & 0 & 0\\
0 & 0 & 0 & 0 & 0 & 0 & 0 & 0 & 0 & 1 & 0 & 0 & 0 & 0\\
0 & 0 & 0 & 0 & 0 & 0 & 0 & 0 & 0 & 0 & \Lambda^{-2\lambda\kappa} & 0 & 0 & 0\\
0 & 0 & 0 & 0 & 0 & 0 & 0 & 0 & 0 & 0 & 0 & 1 & 0 & 0\\
0 & 0 & 0 & 0 & 0 & 0 & 0 & 0 & 0 & 0 & 0 & 0 & \Lambda^{\epsilon} & 0\\
0 & 2k\Lambda^{\epsilon}\ln\Lambda & 0 & 0 & 0 & 0 
& 0 & \frac{\kappa}{2}\Lambda^{\epsilon}\ln\Lambda & 0 & 0 & 0 & 0 
& 0 & \Lambda^{\epsilon} \\
\end{array}\right),
\end{eqnarray} 
hence $P\Lambda^{{\cal N}_{0}}P^{-1}=\Lambda^{{\cal N}}$ 
is uniformly bounded for 
$0<\Lambda<1$ if the condition~(\ref{velocity-condition}) 
and (\ref{velocity-condition'}) hold. 

Thus, there is a unique solution of 
the Fuchsian system~(\ref{fuchsian-eq-gowdy}) which goes to zero as 
$t\rightarrow 0$, and which is analytic in $\theta$ and continuous in $t$. 
Note that $(U,A,\phi,\sigma,\alpha,\eta)$ is a solution of the effective 
evolution equations of the Einstein-matter 
equations (\ref{ce-eta}), (\ref{ce-alpha}), (\ref{ee-u})-(\ref{ee-sigma}) if 
we construct $(U,A,\phi,\sigma,\alpha,\eta)$ 
from (\ref{vtd-sol-eta})-(\ref{vtd-sol-sigma1}) with 
$V=u_{1}$, $B=u_{4}$, $\Phi=u_{7}$, $\Sigma=u_{10}$, 
$\beta=u_{13}$ and $\mu=u_{14}$. This fact follows from equations 
$D(u_{I+2}-t^{\epsilon}u_{I}')=0$, where $I=1,4,7,10$.  

Now, we want to get a constraint condition to ensure that 
the solution obtained above 
is a genuine one to the full Einstein-matter equations. 
Since $D\alpha/\alpha={\cal O}(t^{\epsilon})$, 
\begin{eqnarray}
\frac{\dot{N}}{N}=\frac{\dot{\alpha}}{2\alpha}={\cal O}(t^{\epsilon-1}),
\end{eqnarray} 
then, the right-hand-side of the above equation is integrable. 
From this result, we can put a function $P(t,\theta)$ such that 
\begin{eqnarray}
N\propto \exp{P(t,\theta)}. 
\end{eqnarray} 
This means that $N$ is identically zero 
if we would choose the singular data such that 
$N\rightarrow 0$ as $t\rightarrow 0$, and then, 
the constraint equation~(\ref{ce-eta'}) is 
satisfied. 

Inserting the formal solutions~(\ref{vtd-sol-eta})-(\ref{vtd-sol-sigma1}) 
into the constraint equation~(\ref{N}), we have 
\begin{eqnarray}
N=\eta_{0}'-2kU_{0}'-e^{4U_{0}}(1-2k)h'A_{0}-\frac{\kappa\phi_{0}'}{2}
+e^{2\lambda\phi_{0}}\kappa\omega'\sigma_{0}+\frac{\alpha_{0}'}{2\alpha_{0}}
+{\cal O}(1), 
\end{eqnarray} 
where ${\cal O}(1)$ is some expression which tends 
to zero as $t\rightarrow 0$. 
Thus, the constraint holds iff the singular data satisfy 
\begin{eqnarray}
\label{constraint1}
\eta_{0}'-2kU_{0}'-e^{4U_{0}}(1-2k)h'A_{0}-\frac{\kappa\phi_{0}'}{2}
+e^{2\lambda\phi_{0}}\kappa\omega'\sigma_{0}+\frac{\alpha_{0}'}{2\alpha_{0}}=0.
\end{eqnarray} 

To summarize, we have the following theorem:
\begin{theorem}
\label{avtd1}
Choose data such that conditions (\ref{velocity-condition}), 
(\ref{velocity-condition'}) and (\ref{constraint1}) are satisfied.  
Suppose that 
$\epsilon$ is a positive constant less than 
$\min\{4k,2-4k,-2\lambda\kappa,2+2\lambda\kappa,2K\}$. 
For any choice of the analytic 
singular data $\eta_{0}(\theta)$, $\alpha_{0}(\theta)$, 
$k(\theta)$, $U_{0}(\theta)$, $h(\theta)$, $A_{0}(\theta)$, 
$\kappa(\theta)$, $\phi_{0}(\theta)$, $\omega(\theta)$ 
and $\sigma_{0}(\theta)$, 
the Gowdy symmetric IIA system has a solution of the 
form (\ref{vtd-sol-eta})-(\ref{vtd-sol-sigma1}), where 
$\mu$, $\beta$, $V$, $B$, $\Phi$ and $\Sigma$ 
tend to zero as $t\rightarrow 0$. 
\hfill$\Box$
\end{theorem}
Although the solution given in theorem~\ref{avtd1} 
is generic in the sense that the solution has 
a maximum number 
of free functions, conditions for paradigm-A does not hold since 
$\lambda\kappa<0$, 
i.e. the universe starts with large potential 
and wrong sign of the time derivative of $\phi$. 
To verify the validity of the paradigm-A 
we need to construct a solution allowing a condition 
$\lambda\kappa>0$. Indeed, this problem can be 
overcame as follows. 

If an AVTD solution with $\lambda\kappa>0$ are needed, we replace expansion 
(\ref{vtd-sol-sigma1}) with 
\begin{eqnarray}
\label{vtd-sol-sigma2}
\sigma=\omega(\theta)+t^{\epsilon}\Sigma(t,\theta).
\end{eqnarray} 
In this case, $-2\lambda\kappa$ and $\Lambda^{-2\lambda\kappa}$ 
sitting the 11th line and the 11th row 
in the matrices ${\cal N}$ and $\Lambda^{{\cal N}_{0}}$ 
are replaced by $\epsilon$ and $\Lambda^{\epsilon}$, respectively.  
Also, the constraint condition for the singular data becomes 
\begin{eqnarray}
\label{constraint2}
\eta_{0}'-2kU_{0}'-e^{4U_{0}}(1-2k)h'A_{0}-\frac{\kappa\phi_{0}'}{2}
+\frac{\alpha_{0}'}{2\alpha_{0}}=0.
\end{eqnarray} 
Thus, we have the following 
theorem which is consistent with conditions of paradigm-A. 
\begin{theorem}
\label{avtd2}
Choose data such that conditions (\ref{velocity-condition}), 
(\ref{constraint2}) and 
$\lambda\kappa>-1/2$ are satisfied. 
Suppose that $\epsilon$ 
is a positive constant 
such that $\max\{0,-2\lambda\kappa\}<\epsilon<\min\{4k,2-4k\}$.
For any choice of the analytic 
singular data $\eta_{0}(\theta)$, $\alpha_{0}(\theta)$, 
$k(\theta)$, $U_{0}(\theta)$, $h(\theta)$, $A_{0}(\theta)$, 
$\kappa(\theta)$, $\phi_{0}(\theta)$ and $\omega(\theta)$, 
the Gowdy symmetric IIA system has a solution of the 
form (\ref{vtd-sol-eta})-(\ref{vtd-sol-phi}) and (\ref{vtd-sol-sigma2}), where 
$\mu$, $\beta$, $V$, $B$, $\Phi$ and $\Sigma$ 
tend to zero as $t\rightarrow 0$. 
\hfill$\Box$
\end{theorem}
The positivity of $K$ is automatically satisfied 
when $0<k<1/2$ and $\lambda\kappa>-1/2$ hold. 
Then, a solution to the Gowdy symmetric IIA system allowing 
the initial conditions for paradigm-A has been constructed.
Note that we do not have the maximum number of free functions in this case. 
Thus, the solution given in theorem~\ref{avtd2} is restricted 
than generic one given in theorem~\ref{avtd1}. 
The reason why we do not have the maximum number 
is the existence of dilaton coupling with 
kinetic terms of other fields (the axion field in our case). 
Generically, all fields arising in superstring/M-theory 
couple with the dilaton field. 
Therefore, we may not avoid 
such restriction for solutions to our problem 
unless the dilaton coupling is ignored.

\section{Global existence}
Now, consider the problem (Q2). We will show the following theorem: 
\begin{theorem}
\label{global}
Let $(M,g,\phi,\sigma)$ be the maximal Cauchy development of $C^{\infty}$ 
initial data for the Gowdy symmetric IIA system. 
Suppose that the timelike convergence condition (TCC), which is 
$R_{\mu\nu}W^{\mu}W^{\nu}\geq 0$ for any timelike vector $W^{\mu}$, holds and 
there is a positive constant $\bar{\lambda}$ 
such that $\left|\lambda\right|\leq\bar{\lambda}<1/2$. 
Then, 
$M$ can be covered by compact Cauchy surfaces of constant areal time $t$ 
with each value in the range $(0,\infty)$. 
\end{theorem}

In the first place, we need a local existence theorem for the Gowdy symmetric 
IIA system, which is the Einstein-(minimally coupled) scalar system with a 
positive potential. 
Fortunately, there is no coupling caused by existence of such matter fields 
in the principal part of the PDE system. For this reason, 
since the local existence theorems for vacuum Gowdy (more generically, 
$T^2$-symmetric) spacetimes have been shown~\cite{MV,CP}, 
the same theorem for the Gowdy symmetric 
IIA system can be shown as vacuum case~\cite{FR}. 
Thus, it is enough to verify uniform bounds of functions 
$(\eta,\alpha,U,A,\phi,\sigma)$ and their first and second derivatives to 
prove global existence~\cite{MA}. 
The strategy is similar with the case of $T^2$-symmetric Einstein(-Vlasov) 
system~\cite{AH,ARW,BCIM,IW,WM}.

Let us define 
\begin{eqnarray}
\gamma:=\eta+\frac{1}{2}\ln\alpha.
\end{eqnarray} 
By using $\gamma$ we can rewrite the constraint equations as follows:
\begin{eqnarray}
\label{ce-gamma}
\frac{\dot{\gamma}}{t}={\cal E}-\frac{Q^2}{4}e^{2(\gamma+\lambda\phi-U)},
\end{eqnarray}
\begin{eqnarray}
\label{ce-gamma'}
\frac{\gamma'}{t}=\frac{{\cal F}}{\sqrt{\alpha}},
\end{eqnarray}
\begin{eqnarray}
\label{ce-alpha'}
\dot{\alpha}=-t\alpha Q^{2}e^{2(\gamma+\lambda\phi-U)},
\end{eqnarray}
where 
\begin{eqnarray}
{\cal E}:=\dot{U}^2+\alpha U'^2+\frac{e^{4U}}{4t^2}\left(\dot{A}^2
+\alpha A'^2\right)
+\frac{1}{4}
\left[\dot{\phi}^2+\alpha\phi'^2+e^{2\lambda\phi}\left(\dot{\sigma}^2
+\alpha\sigma'^2\right)\right],
\end{eqnarray} 
and 
\begin{eqnarray}
{\cal F}:=\sqrt{\alpha}\left[2\dot{U}U'+\frac{e^{4U}}{2t^2}\dot{A}A'+
\frac{1}{2}\left(\dot{\phi}\phi'
+e^{2\lambda\phi}\dot{\sigma}\sigma'\right)\right].
\end{eqnarray} 
Define energies for the Gowdy symmetric IIA system
\begin{eqnarray}
E(t):=\int_{S^1}\frac{1}{\sqrt{\alpha}}\left[
{\cal E}
+\frac{1}{4}\alpha Q^2e^{2(\eta+\lambda\phi-U)}
\right]d\theta,
\end{eqnarray} 
and 
\begin{eqnarray}
\tilde{E}(t):=\int_{S^1}\frac{{\cal E}}{\sqrt{\alpha}}
d\theta,
\end{eqnarray} 
In our case, the TCC is as follows:
\begin{eqnarray}
\dot{\phi}^2+e^{2\lambda\phi}\dot{\sigma}^2
\geq\frac{1}{2}\alpha Q^2e^{2(\eta+\lambda\phi-U)}
\end{eqnarray} 

First, we will show energy decay and energy inequalities 
(see lemmas 1 and 3 in \cite{IW}). 
\begin{lemma}
\label{e-decrease}
Suppose the TCC and 
the condition $\left|\lambda\right|\leq\bar{\lambda}<1/2$. Then, 
$E$ and $\tilde{E}$ decrease monotonically along time $t$, that is, 
\begin{eqnarray}
\frac{dE(t)}{dt}<0 \hspace{.5cm}{\rm and}\hspace{.5cm}
\frac{d\tilde{E}(t)}{dt}<0, 
\end{eqnarray} 
and $E$ and $\tilde{E}$ are bounded on 
$(T_{-},T_{+})$, where $0<T_{-}<t_{i}<T_{+}<\infty$. 
Furthermore, there exists numbers, 
$E_{-}$ and $\tilde{E}_{-}$, satisfying 
\begin{eqnarray}
E_{-}=\lim_{t\rightarrow T_{-}}E(t)\hspace{.5cm}{\rm and}\hspace{.5cm}
\tilde{E}_{-}=\lim_{t\rightarrow T_{-}}\tilde{E}(t). 
\end{eqnarray} 
\end{lemma}
{\it Proof}. 
One can calculate directly as follows:
\begin{eqnarray}
\label{dE<0}
\frac{dE(t)}{dt}=-\int_{S^1}\frac{1}{\sqrt{\alpha}t}\left(
2\dot{U}^2+\frac{e^{4U}}{2t^2}\alpha A'^2+\frac{\dot{\phi}^2}{2}
+\frac{e^{2\lambda\phi}\dot{\sigma}^2}{2}
\right)d\theta\leq 0.
\end{eqnarray} 
Thus, $E(t)$ is controlled by $E(t_{i})$ for 
any $t\in[t_{i},T_{+})$. 

The right-hand-side of equation~(\ref{dE<0}) can be controlled by $E$: 
\begin{eqnarray}
\frac{dE}{dt}\geq -\frac{4}{t}E. 
\end{eqnarray} 
For any $t\in(T_{-},t_{i}]$, we have 
\begin{eqnarray}
E(t)\leq E(t_{i})\left(\frac{t_{i}}{t}\right)^4. 
\end{eqnarray} 
Then, $E(t)\leq E(t_{i})\left(\frac{t_{i}}{T_{-}}\right)^4$ on $(T_{-},t_{i})$.
This boundedness and the monotonicity of $E(t)$ assert that $E(t)$ continuously 
extends to $T_{-}$ and then $E_{-}$ exists. 

Next, we show the same results for $\tilde{E(t)}$. 
By direct calculation, we have 
\begin{eqnarray}
\frac{d\tilde{E}(t)}{dt}
=\int_{S^1}&-&\frac{2}{\sqrt{\alpha}t}\left(
\dot{U}^2+\frac{e^{4U}}{4t^2}\alpha A'^2+\frac{\dot{\phi}^2}{4}
+\frac{e^{2\lambda\phi}\dot{\sigma}^2}{4}
\right)\nonumber \\
&+&\frac{\dot{\alpha}}{2\alpha}\left(
\frac{{\cal E}}{\sqrt{\alpha}}+\frac{1}{t\sqrt{\alpha}}\left[
\dot{U}-\lambda\dot{\phi}\right]\right)d\theta.
\end{eqnarray} 
We cannot conclude the monotonicity for $\tilde{E(t)}$ from the above form. 
Now, 
\begin{eqnarray}
-\frac{2\dot{U}^2}{t\sqrt{\alpha}}+\frac{\dot{\alpha}\dot{U}}{2t\alpha\sqrt{\alpha}}
=-\frac{2}{t\sqrt{\alpha}}\left(\dot{U}-\frac{\dot{\alpha}}{8\alpha}\right)^2
+\frac{1}{32t\sqrt{\alpha}}\left(\frac{\dot{\alpha}}{\alpha}\right)^2,
\end{eqnarray} 
and 
\begin{eqnarray}
-\frac{2\dot{\phi}^2}{4t\sqrt{\alpha}}
-\frac{\lambda\dot{\alpha}\dot{\phi}}{2t\alpha\sqrt{\alpha}}
=-\frac{1}{2t\sqrt{\alpha}}
\left(\dot{\phi}+\frac{\lambda\dot{\alpha}}{2\alpha}\right)^2
+\frac{\lambda^2}{8t\sqrt{\alpha}}\left(\frac{\dot{\alpha}}{\alpha}\right)^2.
\end{eqnarray} 
Therefore, 
\begin{eqnarray}
\frac{d\tilde{E}(t)}{dt}
=\int_{S^1}&-&\frac{2}{\sqrt{\alpha}t}\left(
\left(\dot{U}-\frac{\dot{\alpha}}{8\alpha}\right)^2
+\frac{e^{4U}}{4t^2}\alpha A'^2
+\frac{1}{4}\left(\dot{\phi}+\frac{\lambda\dot{\alpha}}{2\alpha}\right)^2
+\frac{e^{2\lambda\phi}\dot{\sigma}^2}{4}
\right)\nonumber \\
&+&\frac{\dot{\alpha}}{2\alpha}\left(
\frac{{\cal E}}{\sqrt{\alpha}}-
\left[\frac{1}{16}+\frac{\lambda^2}{4}\right]
\sqrt{\alpha} Q^2e^{2(\eta+\lambda\phi-U)}\right)d\theta,
\end{eqnarray} 
where equation (\ref{ce-alpha}) has been used. 
By using the TCC and the inequality $\left|\lambda\right|\leq\bar{\lambda}<1/2$, 
we have the conclusion of the monotonic nonincreasing property for 
$\tilde{E(t)}$, 
\begin{eqnarray}
\label{dtildeE<0}
\frac{d\tilde{E}}{dt}\leq\int_{S^1}\frac{C_{\lambda}}{2}
\frac{\dot{\alpha}}{\alpha}\frac{{\cal E}}{\sqrt{\alpha}}d\theta
\leq 0,
\end{eqnarray} 
where $C_{\lambda}<1$ is a positive constant depending only $\lambda$. 

Now, it follows that $\tilde{E(t)}\leq E(t)$ for any time $t$. 
Therefore, one can see that $\tilde{E}(t)$ also extend continuously to 
$T_{-}$ by the monotonicity of it. 
\hfill$\Box$

Next two lemmas will be used to control dynamical parts 
(i.e. $U$, $A$, $\phi$ and $\sigma$) of the system. 
The method of the proof is based on the light cone estimate~\cite{MV,BCIM}. 
\begin{lemma}
\label{lce}
If $\dot{\alpha}\alpha^{-1}$ is bounded, ${\cal E}$ is bounded on 
$(T_{-},T_{+})\times S^1$. 
\end{lemma}
{\it Proof}. 
Differentiating quantities, ${\cal E}$ and ${\cal F}$, along null directions 
$\partial_{\zeta}:=\partial_{t}-\sqrt{\alpha}\partial_{\theta}$ and 
$\partial_{\xi}:=\partial_{t}+\sqrt{\alpha}\partial_{\theta}$, we have 
\begin{eqnarray}
\label{e+m}
\partial_{\zeta}({\cal E}+{\cal F})
=\frac{\dot{\alpha}}{\alpha}({\cal E}+{\cal F})&-&\frac{1}{t}
\left[2\dot{U}^2+\frac{e^{4U}}{2t^2}\alpha A'^2+\frac{1}{2}\dot{\phi}^2
+\frac{1}{2}e^{2\lambda\phi}\dot{\sigma}^2+{\cal F}\right]\nonumber \\
&-&\frac{\dot{\alpha}}{2t\alpha}\left[\dot{U}+\sqrt{\alpha}U'
-\lambda\left(\dot{\phi}
+\sqrt{\alpha}\phi'\right)\right]=:L_{+},
\end{eqnarray} 
and 
\begin{eqnarray}
\label{e-m}
\partial_{\xi}({\cal E}-{\cal F})
=\frac{\dot{\alpha}}{\alpha}({\cal E}-{\cal F})&-&\frac{1}{t}
\left[2\dot{U}^2+\frac{e^{4U}}{2t^2}\alpha A'^2+\frac{1}{2}\dot{\phi}^2
+\frac{1}{2}e^{2\lambda\phi}\dot{\sigma}^2-{\cal F}\right]\nonumber \\
&&\frac{\dot{\alpha}}{2t\alpha}\left[\dot{U}-\sqrt{\alpha}U'
-\lambda\left(\dot{\phi}
-\sqrt{\alpha}\phi'\right)\right]=:L_{-}.
\end{eqnarray} 
Note that 
\begin{eqnarray}
\left(\dot{U}\pm\sqrt{\alpha}U'\right)-\lambda\left(\dot{\phi}
\pm\sqrt{\alpha}\phi'\right)&\leq&
\left(\dot{U}\pm\sqrt{\alpha}U'\right)^2+\lambda^2\left(\dot{\phi}
\pm\sqrt{\alpha}\phi'\right)^2+\frac{1}{2}\nonumber \\
&\leq&
(1+\lambda^2)({\cal E}+{\cal F})+\frac{1}{2}\nonumber \\
&\leq& 2(1+\lambda^2){\cal E}+\frac{1}{2}, 
\end{eqnarray} 
where $\left| {\cal F}\right|\leq {\cal E}$ has been used. 
Thus, 
\begin{eqnarray}
\left|L_{\pm}\right|\leq\left|\frac{\dot{\alpha}}{\alpha}\right|
\left\{2{\cal E}+\frac{C{\cal E}}{t}+\frac{1}{4t}\right\}+\frac{3{\cal E}}{t},
\end{eqnarray} 
where $C$ is a positive constant. 

Consider a point $(t,\theta)\in[t_{i},T_{+})\times S^1$. Integrating 
the both sides of equations (\ref{e+m}) and (\ref{e-m}) 
along null passes, $\partial_{\zeta}$ and 
$\partial_{\xi}$, from points $(t_{i},\theta_{+})$ and 
$(t_{i},\theta_{-})$ to the point $(t,\theta)$, respectively, we have 
\begin{eqnarray}
\int \partial_{\zeta}({\cal E}+{\cal F})d\zeta
={\cal E}(t,\theta)+{\cal F}(t,\theta)
-{\cal E}(t_{i},\theta_{+})-{\cal F}(t_{i},\theta_{+})
=\int L_{+}d\zeta,
\end{eqnarray} 
and 
\begin{eqnarray}
\int \partial_{\xi}({\cal E}-{\cal F})d\xi
={\cal E}(t,\theta)-{\cal F}(t,\theta)
-{\cal E}(t_{i},\theta_{-})+{\cal F}(t_{i},\theta_{-})
=\int L_{-}d\xi.
\end{eqnarray} 
Adding these equations and using the inequality 
$\left|{\cal F}\right|\leq {\cal E}$, 
\begin{eqnarray}
\label{ell}
{\cal E}(t,\theta)\leq {\cal E}(t_{i},\theta_{+})+{\cal E}(t_{i},\theta_{-})+
\frac{1}{2}\left[\int \left|L_{+}\right|d\zeta
+\int \left|L_{-}\right|d\xi\right]. 
\end{eqnarray} 
Taking supremums over all values of the space coordinate $\theta$ on the both 
sides of the inequality (\ref{ell}), we have 
\begin{eqnarray}
\label{ecc}
\sup_{S^1}{\cal E}(t,\theta)&\leq&
 2\sup_{S^1}{\cal E}(t_{i},\theta)
+\int^{t}_{t_{i}}\left[\left|\frac{\dot{\alpha}}{\alpha}\right|
\left\{2\sup_{S^1}{\cal E}\left(1+\frac{C}{s}\right)+\frac{1}{4s}\right\}
+\frac{3}{s}\sup_{S^1}{\cal E}\right]ds\nonumber \\
&=&C_{1}(t)+\int^{t}_{t_{i}}C_{2}(s)\sup_{S^1}{\cal E}(s,\theta)ds,
\end{eqnarray} 
where $C_{i}(t)$ are bounded and positive functions of $t$. 
We now apply Gronwall's lemma to this inequality (\ref{ecc}), 
we have boundedness 
for ${\cal E}$ on $[t_{i},T_{+})\times S^1$. 
We can apply the same argument for $t\in(T_{-},t_{i}]\times S^1$, and then 
we have the conclusion of this lemma.
\hfill$\Box$
\begin{lemma}
\label{lce2}
Let us define 
\begin{eqnarray}
\tilde{{\cal E}}:=\ddot{U}^2+\alpha\dot{U}'^2
+\frac{e^{4U}}{4t^2}\left(\ddot{A}^2+\alpha\dot{A}'^2\right)
+\frac{1}{4}\left[\ddot{\phi}^2+\alpha\dot{\phi}'^2
+e^{2\lambda\phi}\left(\ddot{\sigma}^2
+\alpha\dot{\sigma}'^2\right)\right],
\end{eqnarray} 
and 
\begin{eqnarray}
\tilde{{\cal F}}:=\sqrt{\alpha}\left[2\ddot{U}\dot{U}'
+\frac{e^{4U}}{2t^2}\ddot{A}\dot{A}'+
\frac{1}{2}\left(\ddot{\phi}\dot{\phi}'
+e^{2\lambda\phi}\ddot{\sigma}\dot{\sigma}'\right)\right].
\end{eqnarray} 
If all functions and their first derivative, 
$\dot{\alpha}'$ and $\ddot{\alpha}$ are bounded, $\tilde{{\cal E}}$ 
is bounded on 
$(T_{-},T_{+})\times S^1$. 
\end{lemma}
{\it Proof}. 
Taking time derivative of the wave equations~(\ref{ee-u})-(\ref{ee-sigma}) for 
$U$, $A$, $\phi$ and $\sigma$, we have wave equations for $\dot{U}$, 
$\dot{A}$, $\dot{\phi}$ and $\dot{\sigma}$.  Now, $\tilde{{\cal E}}$ and 
$\tilde{{\cal F}}$ satisfy equations of the form 
\begin{eqnarray}
\partial_{\zeta}(\tilde{{\cal E}}+\tilde{{\cal F}})=\tilde{L}_{+}
\hspace{.5cm}{\rm and}\hspace{.5cm}
\partial_{\xi}(\tilde{{\cal E}}-\tilde{{\cal F}})=\tilde{L}_{-},
\end{eqnarray}
where $\tilde{L}_{\pm}$ involve nothing but controlled quantities, together 
with terms quadratic in $\ddot{U}$, $\dot{U}'$, $\ddot{A}$, $\dot{A}'$, 
$\ddot{\phi}$, $\dot{\phi}'$, $\ddot{\sigma}$ and $\dot{\sigma}'$. 
Now, we can repeat the light cone argument and then, we have boundedness for 
$\tilde{{\cal E}}$ on 
$(T_{-},T_{+})\times S^1$. 
\hfill$\Box$

\subsection{Past direction}
Further estimates are given in the each case of past and future directions, 
separately. First, consider the past direction. 

\begin{lemma}
\label{max-min-gamma-u-phi}
For any $t$, the function $\gamma$ satisfies the following condition, 
\begin{eqnarray}
\max_{S^1}\gamma(t,\theta)-\min_{S^1}\gamma(t,\theta)\leq tE(t).
\end{eqnarray}
Furthermore, for any $t\in (T_{-},t_{i}]$, the functions $U$ and $\phi$ satisfy 
the following conditions, 
\begin{eqnarray}
\max_{S^1}U(t,\theta)-\min_{S^1}U(t,\theta)\leq CE^{1/2}(t), 
\end{eqnarray}
and 
\begin{eqnarray}
\max_{S^1}\phi(t,\theta)-\min_{S^1}\phi(t,\theta)\leq CE^{1/2}(t). 
\end{eqnarray}
\end{lemma}
{\it Proof}. (cf. Step 1 of Section 5 in \cite{AH}). 
For any $\theta_{1},\theta_{2}\in S^1$, we have 
\begin{eqnarray}
\left|\gamma(t,\theta_{2})-\gamma(t,\theta_{1})\right|=
\left|\int^{\theta_{2}}_{\theta_{1}}\gamma'd\theta\right|\leq
\int^{\theta_{2}}_{\theta_{1}}\left|\gamma'\right| d\theta \leq
\int^{\theta_{2}}_{\theta_{1}}\frac{t{\cal E}}{\sqrt{\alpha}}d\theta \leq
t\tilde{E}(t)\leq tE(t),
\end{eqnarray}
where equation (\ref{ce-gamma'}) 
and the fact $\left|{\cal F}\right|\leq{\cal E}$ have been used. 
Since $\theta_{1}$ and $\theta_{2}$ are arbitrary, the first conclusion follows. 

Similarly, for any $\theta_{1},\theta_{2}\in S^1$ and any $t\in (T_{-},t_{i}]$, 
we have
\begin{eqnarray}
\left|U(t,\theta_{2})-U(t,\theta_{1})\right|&=&
\left|\int^{\theta_{2}}_{\theta_{1}}U'd\theta\right|\nonumber \\
&\leq&
\left(\int^{\theta_{2}}_{\theta_{1}}\frac{d\theta}{\sqrt{\alpha}}\right)^{1/2}
\left(\int^{\theta_{2}}_{\theta_{1}}\sqrt{\alpha}U'^2d\theta\right)^{1/2}
\nonumber \\
&\leq&
\left(\int^{\theta_{2}}_{\theta_{1}}\frac{d\theta}{\sqrt{\alpha(t_{i})}}\right)^{1/2}
\tilde{E}(t)^{1/2}
\nonumber \\
&\leq&
CE(t)^{1/2},
\end{eqnarray}
where the H\"older inequality and the monotonicity of $\alpha$ have been used. 

The proof for $\phi$ is used the same argument.
\hfill$\Box$

\begin{lemma}
The function $\gamma$ is bounded from above on 
$(T_{-},t_{i}]\times S^1$.
\end{lemma}
{\it Proof}. (cf. lemma 4 in \cite{IW}). 
Note that 
\begin{eqnarray}
\label{tcc-initial}
\dot{\phi}(t_{i})^2+e^{2\lambda\phi(t_{i})}\dot{\sigma}(t_{i})^2
\geq\frac{1}{2}\alpha(t_{i}) Q^2e^{2[\eta(t_{i})+\lambda\phi(t_{i})-U(t_{i})]}
>0,
\end{eqnarray}
since regular initial data at $t=t_{i}$ are supposed. 
This means $\tilde{E}(t_{i})>0$. 
From equation (\ref{dtildeE<0}), we have 
\begin{eqnarray}
\frac{d\tilde{E}}{dt}\leq\int_{S^1}\frac{C_{\lambda}}{2}
\frac{\dot{\alpha}}{\alpha}\frac{{\cal E}}{\sqrt{\alpha}}d\theta
=-\frac{C_{\lambda}Q^2}{2}\int_{S^1}te^{2(\gamma+\lambda\phi-U)}
\frac{{\cal E}}{\sqrt{\alpha}}d\theta,
\end{eqnarray}
where $C_{\lambda}<1$ is a positive constant 
depending on only the coupling constant $\lambda$. 
Suppose $\lambda\geq 0$. 
Integrating this inequality from $t_{i}$ to $t$ ($0<t<t_{i}$), 
\begin{eqnarray}
\label{tildeE>tildeEi}
\tilde{E}(t)&\geq&\tilde{E}(t_{i})
+\frac{C_{\lambda}Q^2}{2}\int^{t_{i}}_{t}\left(\int_{S^1}se^{2(\gamma+\lambda\phi-U)}
\frac{{\cal E}}{\sqrt{\alpha}}d\theta\right)ds\nonumber \\
&\geq&\tilde{E}(t_{i})
+\frac{C_{\lambda}Q^2}{2}\int^{t_{i}}_{t}
s\exp\left[2\left(\min_{S^1}\gamma+\lambda\min_{S^1}\phi-\max_{S^1}U\right)\right]
\tilde{E}(s)ds\nonumber \\
&\geq&\tilde{E}(t_{i})\left(1+
\frac{C_{\lambda}Q^2}{2}\int^{t_{i}}_{t}
s\exp\left[2\left(\min_{S^1}\gamma+\lambda\min_{S^1}\phi-\max_{S^1}U\right)\right]
ds\right),
\end{eqnarray}
where the monotonicity of $\tilde{E}$ has been used. 
From lemma~\ref{max-min-gamma-u-phi}, 
\begin{eqnarray}
\label{minminmax}
&&
\min_{S^1}\gamma+\lambda\min_{S^1}\phi-\max_{S^1}U\nonumber \\
&\geq&
\max_{S^1}\gamma+\lambda\max_{S^1}\phi-\min_{S^1}U
-\left(tE(t)+(C_1\lambda-C_2)E(t)^{1/2}\right)\nonumber \\
&\geq&
\max_{S^1}\gamma+\lambda\max_{S^1}\phi-\min_{S^1}U
-\left(t_{i}E(T_{-})+(C_1\lambda-C_2)E(\tau)^{1/2}\right),
\end{eqnarray}
where $C_{1}$ and $C_{2}$ are positive constants, 
$\tau=t_{i}$ if $C_1\lambda-C_2< 0$ 
and $\tau=T_{-}$ if $C_1\lambda-C_2\geq 0$.
Thus, we have 
\begin{eqnarray}
\tilde{E}(t)\geq\tilde{E}(t_{i})\left(1+
\frac{C_{\lambda}Q^2}{2}
e^{-2\left(t_{i}E(T_{-})+(C_1\lambda-C_2)E(\tau)^{1/2}\right)}
\int^{t_{i}}_{t}
se^{2\left(\gamma+\lambda\phi-U\right)}
ds\right),
\end{eqnarray}
and then, 
\begin{eqnarray}
\label{te<}
\int^{t_{i}}_{t}
se^{2\left(\gamma+\lambda\phi-U\right)}ds\leq
\frac{2}{C_{\lambda}Q^2}
e^{2\left(t_{i}E(T_{-})+(C_1\lambda-C_2)E(\tau)^{1/2}\right)}
\left(\frac{\tilde{E}(T_{-})}{\tilde{E}(t_{i})}-1\right),
\end{eqnarray}
where the condition (\ref{tcc-initial}) has been used. 
When one consider the case of $\lambda<0$, we have the same results by exchanging 
$\max_{S^1}\phi$ and $\min_{S^1}\phi$ in inequalities 
(\ref{tildeE>tildeEi}) and (\ref{minminmax}).

Now, integrating equation (\ref{ce-gamma}), we have 
\begin{eqnarray}
\gamma(t,\theta)&=&\gamma(t_{i},\theta)-\int^{t_{i}}_{t}\left[s{\cal E}
-\frac{sQ^2}{4}e^{2\left(\gamma+\lambda\phi-U\right)}\right]ds\nonumber \\
&\leq&
\gamma(t_{i},\theta)+\frac{Q^2}{4}\int^{t_{i}}_{t}
se^{2\left(\gamma+\lambda\phi-U\right)}ds\nonumber \\
&\leq&
\max_{S^1}\gamma(t_{i},\theta)+\frac{1}{C_{\lambda}2}
e^{2\left(t_{i}E(T_{-})+(C_1\lambda-C_2)E(\tau)^{1/2}\right)}
\left(\frac{\tilde{E}(T_{-})}{\tilde{E}(t_{i})}-1\right).
\end{eqnarray}
Thus, the boundedness of $\gamma$ from above has been shown. 
\hfill$\Box$

\begin{lemma}
\label{n<1/2}
For any numbers $a$ and $b$, and for $n\leq\frac{1}{2}$, 
$\alpha^{n}e^{2\eta+a\phi-bU}$ is bounded on 
$(T_{-},t_{i}]\times S^1$.
\end{lemma}
{\it Proof}. (cf. lemma 5 in \cite{WM}).
\begin{eqnarray}
&&\partial_{t}\left(t^{k}\alpha^{n}e^{2\eta+a\phi-bU}\right)
\nonumber \\
&=&
\left(\frac{k}{t}
+\frac{n\dot{\alpha}}{\alpha}+2\dot{\eta}+a\dot{\phi}-b\dot{U}
\right)t^{k}\alpha^{n}e^{2\eta+a\phi-bU}\nonumber \\
&=&
\left[
2t\left(\dot{U}-\frac{b}{4t}\right)^2
+\frac{t}{2}\left(\dot{\phi}+\frac{a}{t}\right)^2+
2\alpha U'^2+\frac{e^{4U}}{2t^2}\left(\dot{A}^2+\alpha A'^2\right)
\right.
\nonumber \\
&&
\left.
+\frac{1}{2}\left(\alpha\phi'^2+e^{2\lambda\phi}(\dot{\sigma}^2+\alpha\sigma'^2)
\right)+\left(\frac{1}{2}-n\right)t\alpha Q^2e^{2(\eta+\lambda\phi-U)}\right]
t^{k}\alpha^{n}e^{2\eta+a\phi-bU}
\nonumber \\
&\geq&0, 
\end{eqnarray} 
where we have chosen $8k=4a^2+b^2$. Then, we have 
\begin{eqnarray}
\alpha(t,\theta)^{n}e^{2\eta(t,\theta)+a\phi(t,\theta)-bU(t,\theta)}
\leq\left(\frac{t_{i}}{T_{-}}\right)^k
\alpha(t_{i},\theta)^{n}
e^{2\eta(t_{i},\theta)+a\phi(t_{i},\theta)-bU(t_{i},\theta)},
\end{eqnarray} 
on $(T_{-},t_{i}]\times S^1$. 
\hfill$\Box$

\begin{lemma}
\label{min-alpha}
$\alpha$ is bounded on $(T_{-},t_{i}]\times S^1$. 
\end{lemma}
{\it Proof}. 
Integrating the constraint equation (\ref{ce-alpha}), we have 
\begin{eqnarray}
-\int^{t_{i}}_{t}\frac{\dot{\alpha}}{\alpha}ds=
\ln\alpha(t)-\ln\alpha(t_{i})=
Q^2\int^{t_{i}}_{t}se^{2(\gamma+\lambda\phi-U)}ds,
\end{eqnarray} 
for $t\in(T_{-},t_{i}]$. By using inequality (\ref{te<}), 
we have boundedness of $\ln\alpha$ from above. As a result, $0<\alpha$ is also 
bounded. 
\hfill$\Box$

\begin{lemma}
\label{e-gamma}
For any numbers $a$ and $b$, $e^{\gamma+a\phi-bU}
(=\sqrt{\alpha}e^{\eta+a\phi-bU})$ is bounded 
on $(T_{-},t_{i}]\times S^1$. 
\end{lemma}
{\it Proof}. 
We have already a result that $e^{2\eta+a\phi-bU}$ is bounded 
on $(T_{-},t_{i}]\times S^1$ (lemma~\ref{n<1/2}).  
Combining this and lemma~\ref{min-alpha}, 
the boundedness of 
$e^{\gamma+a\phi-bU}$ on $(T_{-},t_{i}]\times S^1$ follows directly. 
\hfill$\Box$

\begin{corollary}
\label{ln-alpha-past}
$\dot{\alpha}\alpha^{-1}=\partial_{t}(\ln\alpha)$ is bounded on 
$(T_{-},t_{i}]\times S^1$. Thus, 
$\ln\alpha$ and $\dot{\alpha}$ are as well.
\end{corollary}
{\it Proof}. 
Boundedness of $\alpha e^{2(\eta+a\phi-bU)}$ 
is obtained by lemma~\ref{e-gamma}. 
From the constraint equation (\ref{ce-alpha'}), we have 
$\dot{\alpha}\alpha^{-1}=-t\alpha Q^2e^{2(\lambda\phi+\eta-U)}$. 
If we set $a=\lambda$ and $b=1$, the boundedness of the right-hand-side of 
that equation is obtained. Thus, the conclusion of this lemma is shown. 
\hfill$\Box$

\begin{lemma}
\label{1st-past}
The functions $U$, $A$, $\phi$, $\sigma$ and their first derivatives 
are bounded on $(T_{-},t_{i}]\times S^1$. 
\end{lemma}
{\it Proof}. 
From lemma~\ref{lce} and corollary~\ref{ln-alpha-past}, 
we have the boundedness for ${\cal E}$ on $(T_{-},t_{i}]\times S^1$. 
Then, $\left|\dot{U}\right|$, $\left|U'\right|$, $\left|\dot{\phi}\right|$, 
$\left|\phi'\right|$, $\left|\left(e^{2U}/2t\right)\dot{A}\right|$, 
$\left|\left(e^{2U}/2t\right)A'\right|$, 
$\left|e^{\lambda\phi}\dot{\sigma}\right|$ 
and  $\left|e^{\lambda\phi}\sigma'\right|$ 
are bounded for all $t\in(T_{-},T_{+})$. 
Once the boundedness on the first derivative of $U$ and $\phi$ is obtained, 
it follows that $U$ and $\phi$ are bounded for all $t\in(T_{-},T_{+})$. 
Then, we have bounds on $\dot{A}$, $A'$, $\dot{\sigma}$ and $\sigma'$, and 
consequently on $A$ and $\sigma$.
\hfill$\Box$

\begin{lemma}
\label{alpha-gamma-eta}
The functions $\alpha'$, $\dot{\alpha}'$ and $\ddot{\alpha}$ are 
bounded on $(T_{-},t_{i}]\times S^1$. Also, $\eta$, $\dot{\eta}$ and 
$\eta'$ are as well.
\end{lemma}
{\it Proof}. (cf. Step 3 of Section 6 in \cite{BCIM}). 
From the constraint equations (\ref{ce-gamma}) and (\ref{ce-gamma'}), 
we have boundedness for $\dot{\gamma}$ and $\gamma'$ directly. Then, 
$\gamma$ is controlled. 
Differentiating both side of equation (\ref{ce-alpha'}) with respect to 
$\theta$, we have 
\begin{eqnarray}
\label{alpha-dot'}
\dot{\alpha}'=\alpha'\left(-tQ^2e^{2(\gamma+\lambda\phi-U)}\right)
-2tQ^2e^{2(\gamma+\lambda\phi-U)}\alpha\left(\gamma'+\lambda\phi'-U'\right). 
\end{eqnarray} 
Then, we have boundedness for $\alpha'$ by integrating this differential 
equation for $\alpha'$ in time since the coefficient of $\alpha'$ and the 
second term in the right-hand-side of the equation (\ref{alpha-dot'}) 
are controlled. 
Thus, we have that $\eta$, $\dot{\eta}$ and 
$\eta'$ is bounded. 

The boundedness of $\dot{\alpha}'$ is obtained immediately from 
(\ref{alpha-dot'}). Now, differentiating both side of equation 
(\ref{ce-alpha'}) with respect to 
$t$, we have 
\begin{eqnarray}
\ddot{\alpha}=-Q^2\alpha e^{2(\eta+\lambda\phi-U)}
\left[\alpha+2t\dot{\alpha}+2t\alpha\left(\dot{\eta}
+\lambda\dot{\phi}-\dot{U}\right)\right], 
\end{eqnarray} 
which implies that $\ddot{\alpha}$ is bounded.  
\hfill$\Box$

\begin{lemma}
\label{2nd-past}
The second derivatives of $U$, $A$, $\phi$ and $\sigma$ 
are bounded on $(T_{-},t_{i}]\times S^1$. 
\end{lemma}
{\it Proof}. 
By lemma~\ref{lce2} we have the boundedness for $\tilde{{\cal E}}$ 
on $(T_{-},t_{i}]\times S^1$. 
Then, we have uniform bounds on $\ddot{U}$, $\dot{U}'$, $\ddot{A}$, 
$\dot{A}'$, $\ddot{\phi}$, $\dot{\phi}'$, $\ddot{\sigma}$ and 
$\dot{\sigma}'$.  Bounds on $U''$, $A''$, $\phi''$ and $\sigma''$ follows from 
the wave equations (\ref{ee-u})-(\ref{ee-sigma}) directly. 
\hfill$\Box$

\begin{lemma}
\label{eta''}
$\alpha''$, $\ddot{\eta}$, $\dot{\eta}'$ and $\eta''$ are 
bounded on $(T_{-},t_{i}]\times S^1$. 
\end{lemma}
{\it Proof}. 
By taking the time derivative of (\ref{ce-gamma}) and (\ref{ce-gamma'}), 
we have bounds on $\ddot{\gamma}$ and $\dot{\gamma}'$. 
Then, bounds on $\ddot{\eta}$ and $\dot{\eta}'$ are obtained 
by the definition of $\gamma$. 
Also, by taking the $\theta$ derivative of (\ref{ce-gamma'}), 
we have bounds on $\gamma''$. 
The boundedness for $\alpha''$ follows from the same argument 
in the proof of lemma~\ref{alpha-gamma-eta}. 
That is, 
differentiating both side of equation~(\ref{alpha-dot'}) with respect to 
$\theta$, we have
\begin{eqnarray}
\label{alpha-dot''}
\dot{\alpha}''&=&\alpha''\left(-tQ^2e^{2(\gamma+\lambda\phi-U)}\right)
-4tQ^2\alpha'(\gamma'+\lambda\phi'-U')e^{2(\gamma+\lambda\phi-U)}\\ \nonumber 
&-&2tQ^2e^{2(\gamma+\lambda\phi-U)}\alpha\left[\gamma''+\lambda\phi''
-U''+2\left(\gamma'+\lambda\phi'-U'\right)^2\right]. 
\end{eqnarray} 
Therefore, we have boundedness for $\alpha''$ by integrating this differential 
equation for $\alpha''$ in time since the coefficient of $\alpha''$ and the 
second and third terms in the right-hand-side of the equation 
(\ref{alpha-dot''}) are bounded as shown already.  
Then, $\eta''$ is 
bounded by using the wave equation~(\ref{ee-eta}).
\hfill$\Box$

\subsection{Future direction}
Now, consider the future direction. 
We have already a monotonic decreasing property of $E(t)$ 
along increasing $t$, $dE/dt<0$ (lemma~\ref{e-decrease}). 
Therefore, for any $t\in[t_{i},T_{+})$, 
\begin{eqnarray}
\label{e<ei}
E(t)\leq E(t_{i}). 
\end{eqnarray} 

Proofs of the following two lemmas are similar with the argument in 
Step 1 of Section 5 in \cite{AH}. 
\begin{lemma}
\label{int-alpha-1/2}
$\int_{\theta_{1}}^{\theta_{2}}\alpha^{-1/2}d\theta$ 
is bounded on $[t_{i},T_{+})$.
\end{lemma}
{\it Proof}. 
The constraint equation (\ref{ce-alpha}) can be written as 
\begin{eqnarray}
\partial_{t}(\alpha^{-1/2})
=\frac{t}{2}\sqrt{\alpha}Q^2e^{2(\eta+\lambda\phi-U)}.
\end{eqnarray} 
Then, 
\begin{eqnarray}
\alpha^{-1/2}(t,\theta)-\alpha^{-1/2}(t_{i},\theta)=
\frac{Q^2}{2}\int^{t}_{t_{i}}s\sqrt{\alpha}e^{2(\eta+\lambda\phi-U)}ds,
\end{eqnarray} 
for ant $t\in[t_{i},T_{+})$. Integrating this equation 
from $\theta_{1}$ to $\theta_{2}$ in $S^1$, we have 
\begin{eqnarray}
\int_{\theta_{1}}^{\theta_{2}}\alpha^{-1/2}(t,\theta)d\theta&=&
\frac{Q^2}{2}\int^{t}_{t_{i}}s\int_{\theta_{1}}^{\theta_{2}}
\sqrt{\alpha}e^{2(\eta+\lambda\phi-U)}d\theta ds
+\int_{\theta_{1}}^{\theta_{2}}\alpha^{-1/2}(t_{i},\theta)d\theta
\nonumber \\
&\leq&
\frac{Q^2}{2}E(t_{i})\int^{t}_{t_{i}}sds
+2\pi\sup_{S^1}\alpha^{-1/2}(t_{i},\theta)
\nonumber \\
&\leq&
\frac{Q^2}{4}E(t_{i})(t^2-t_{i}^2)+C,
\end{eqnarray} 
where (\ref{e<ei}) has been used. 
\hfill$\Box$
\begin{lemma}
\label{u-phi-bound}
The functions $U$ and $\phi$ are bounded on $[t_{i},T_{+})\times S^1$.
\end{lemma}

{\it Proof}. 
For any $\theta_{1}$, $\theta_{2}$ and for each $t\in[t_{i},T_{+})$, 
\begin{eqnarray}
\label{u-u<c}
\left|U(t,\theta_{2})-U(t,\theta_{1})\right|
&=&\left|\int_{\theta_{1}}^{\theta_{2}}U'd\theta\right|\nonumber \\
&\leq& \left(\int_{\theta_{1}}^{\theta_{2}}\alpha^{-1/2}d\theta\right)^{1/2}
\left(\int_{\theta_{1}}^{\theta_{2}}\alpha^{1/2}U'^2d\theta\right)^{1/2}
\nonumber \\
&\leq& C(t),  
\end{eqnarray} 
where the H\"older inequality, energy bound~(\ref{e<ei}) and 
lemma~\ref{int-alpha-1/2} have been used. 

Now, 
\begin{eqnarray}
\left|\int_{S^1}U(t,\theta)d\theta\right|&=&
\left|\int_{t_{i}}^{t}\int_{S^1}\dot{U}(t,\theta)d\theta ds+C\right|
\nonumber \\
&\leq &
\int_{t_{i}}^{t}\int_{S^1}\left|\dot{U}(t,\theta)\right|d\theta ds
+\left|C\right|
\nonumber \\
&\leq &
\int_{t_{i}}^{t}\left(\int_{S^1}\alpha^{1/2}d\theta\right)^{1/2}
\left(\int_{S^1}\alpha^{-1/2}\dot{U}^2d\theta\right)^{1/2}ds+\left|C\right|,  
\end{eqnarray} 
where the H\"older inequality has been used. Since 
$\alpha$ is monotonically decreasing function along increasing time $t$, 
the right-hand-side of the above inequality can be bounded. Thus, 
\begin{eqnarray}
\label{u<c}
\left|\int_{S^1}U(t,\theta)d\theta\right|\leq C(t),
\end{eqnarray} 
for some uniformly bounded function $C(t)$.

Finally, we obtain a uniform bound on $U$. We have the following identity:
\begin{eqnarray}
2\pi\max_{S^1}U(t,\theta)=\int_{S^1}U(t,\theta)d\theta
+\int_{S^1}\left(\max_{S^1}U(t,\theta)-U(t,\theta)\right)d\theta. 
\end{eqnarray} 
The right-hand-side of this identity is bounded from above 
since (\ref{u-u<c}) and (\ref{u<c}) hold. . 
For $\min_{S^1}U(t,\theta)$, one can use the same argument and then, 
$\min_{S^1}U(t,\theta)$ is bounded from below. Thus, $U$ is uniformly bounded 
on $[t_{i},T_{+})\times S^1$. 

We can obtain uniform boundedness for $\phi$ by replacing $U$ with $\phi$ 
in the above argument. 
\hfill$\Box$

\begin{lemma}
\label{gamma-bound}
The functions $\gamma$ is bounded on $[t_{i},T_{+})\times S^1$.
\end{lemma}
{\it Proof}. (cf. Step 1 of Section 6 in \cite{BCIM}). 
From the constraint equation (\ref{ce-gamma}) for $\gamma$, 
we have two inequalities: 
\begin{eqnarray}
\label{gamma<te}
\dot{\gamma}\leq te,
\end{eqnarray} 
and 
\begin{eqnarray}
\label{gamma>tqe}
\dot{\gamma}\geq -\frac{1}{4}tQ^2e^{2(\gamma+\lambda\phi-U)}.
\end{eqnarray} 

From the inequality (\ref{gamma<te}), we have 
\begin{eqnarray}
\int_{S^1}\gamma(t,\theta)d\theta-\int_{S^1}\gamma(t_{i},\theta)d\theta
&=&\int^{t}_{t_{i}}\frac{d}{ds}\left(\int_{S^1}\gamma(s,\theta)d\theta\right)ds
\nonumber \\
&\leq&\sup_{S^1}\sqrt{\alpha}(t_{i},\theta)\int^{t}_{t_{i}}sE(s)ds
\nonumber \\
&\leq&
C\int^{t}_{t_{i}}sE(t_{i})ds
\nonumber \\
&=&\frac{CE(t_{i})}{2}(t^2-t^2_{i}),
\end{eqnarray} 
which controls $\int_{S^1}\gamma(t,\theta)d\theta$ from above. 
Now, we have the following identity: 
\begin{eqnarray}
\label{gamma=max-gamma}
\int_{S^1}\gamma(t,\theta)d\theta=2\pi\max_{S^1}\gamma
+\int_{S^1}\left(\gamma-\max_{S^1}\gamma\right)d\theta.
\end{eqnarray} 
By the equation (\ref{ce-gamma'}) of $\gamma$ and a basic inequality, we have 
\begin{eqnarray}
\int_{S^1}\left|\gamma'\right|d\theta\leq tE(t_{i}).
\end{eqnarray} 
Then, 
\begin{eqnarray}
\label{gamma-gamma<te}
\left|\gamma(t,\theta_{2})-\gamma(t,\theta_{1})\right|
=\left|\int_{\theta_{1}}^{\theta_{2}}\gamma'd\theta\right|
\leq \int_{\theta_{1}}^{\theta_{2}}\left|\gamma'\right|d\theta\leq
\int_{S^1}\left|\gamma'\right|d\theta
\leq tE(t_{i}).
\end{eqnarray} 
Therefore, combining (\ref{gamma=max-gamma}) and (\ref{gamma-gamma<te}), 
we have the upper bound 
for $\gamma$:
\begin{eqnarray}
\label{max-gamma<c}
\max_{S^1}\gamma\leq C(t),
\end{eqnarray} 
where $C(t)$ is a bounded function of $t\in[t_{i},T_{+})$. 

From the inequalities (\ref{gamma>tqe}) and (\ref{max-gamma<c}), 
and lemma~\ref{u-phi-bound}, 
if the coupling constant $\lambda$ is non-negative, 
\begin{eqnarray}
\dot{\gamma}\geq -\frac{1}{4}tQ^2e^{2(\gamma+\lambda\phi-U)}
\geq Ct\exp\left[2\left(\max_{S^1}\gamma+\lambda\max_{S^1}\phi
-\min_{S^1}U\right)\right]
\geq Cte^{c(t)}, 
\end{eqnarray} 
for some bounded function $c(t)$ of $t\in[t_{i},T_{+})$ and $C<0$. 
If $\lambda$ is negative, $\max_{S^1}\phi$ is replaced by $\min_{S^1}\phi$. 
Thus, $\dot{\gamma}$ is controlled into the future, so we have upper and 
lower bounds for $\gamma$ on $[t_{i},T_{+})\times S^1$.
\hfill$\Box$

\begin{corollary}
\label{ln-alpha-future}
$\dot{\alpha}\alpha^{-1}$ {\rm (}hence $\ln\alpha$ and $\alpha${\rm )}, 
$\eta$ and $\dot{\alpha}$ are 
bounded on $[t_{i},T_{+})\times S^1$.
\end{corollary}
{\it Proof}. 
The constraint equation (\ref{ce-alpha'}) can be written as 
\begin{eqnarray}
\frac{\dot{\alpha}}{\alpha}=-tQ^2e^{2(\gamma+\lambda\phi-U)}. 
\end{eqnarray} 
With boundedness of $\gamma$ (lemma~\ref{gamma-bound}), $\phi$ and $U$ 
(lemma~\ref{u-phi-bound}), 
$\dot{\alpha}\alpha^{-1}=\partial_{t}\left(\ln\alpha\right)$ is bounded on 
$[t_{i},T_{+})\times S^1$. As immediate results, $\ln\alpha$ and $\alpha$ are 
bounded on $[t_{i},T_{+})\times S^1$. 
Since $\eta=\gamma-\frac{1}{2}\ln\alpha$, $\eta$ is bounded 
on $[t_{i},T_{+})\times S^1$.
Using these results to the constraint equation (\ref{ce-alpha'}), 
we have a conclusion that 
$\dot{\alpha}$ is bounded on $[t_{i},T_{+})\times S^1$. 
\hfill$\Box$

Once the boundedness of $\dot{\alpha}\alpha^{-1}$ has been obtained, 
the following arguments are similar with ones of the past direction 
because key lemmas (lemma~\ref{lce} and lemma~\ref{lce2}) can be used 
and the arguments do not depend on time directions. 

\begin{lemma}
The functions $U$, $A$, $\phi$, $\sigma$ and their first derivatives 
are bounded on $[t_{i},T_{+})\times S^1$. 
\end{lemma}
{\it Proof}. 
From lemma~\ref{lce} and corollary~\ref{ln-alpha-future}, 
we have the boundedness for ${\cal E}$ on $[t_{i},T_{+})\times S^1$. 
The proof is the same with one of 
lemma~\ref{1st-past}. 
\hfill$\Box$

\begin{lemma}
The functions $\dot{\eta}$, $\eta'$, $\alpha'$, $\dot{\alpha}'$ and 
$\ddot{\alpha}$ are bounded on $[t_{i},T_{+})\times S^1$.
\end{lemma}
{\it Proof}. 
Since the constraint equation (\ref{ce-eta}) is described in terms of 
bounded functions and $t$, we have bounds on $\dot{\eta}$.
From the constraint equation (\ref{ce-gamma'}), we have bounds on $\gamma'$ 
and then, boundedness for $\alpha'$, $\dot{\alpha}'$ and 
$\ddot{\alpha}$ can be obtained by the same argument with 
the proof of lemma~\ref{alpha-gamma-eta}. 
Combining this result, boundedness of $\gamma'$ and 
definition of $\gamma$, we have that $\eta'$ is bounded. 
\hfill$\Box$

\begin{lemma}
\label{2nd-future}
The second derivatives of $U$, $A$, $\phi$ and $\sigma$ 
are bounded on $[t_{i},T_{+})\times S^1$. 
\end{lemma}
{\it Proof}. 
By lemma~\ref{lce2} we have the boundedness for $\tilde{{\cal E}}$ on 
$[t_{i},T_{+})\times S^1$. 
The proof is the same with one of lemma~\ref{2nd-past}. 
\hfill$\Box$

\begin{lemma}
$\alpha''$, $\ddot{\eta}$, $\dot{\eta}'$ and $\eta''$ are 
bounded on $[t_{i},T_{+})\times S^1$. 
\end{lemma}
{\it Proof}. 
The argument is the same to lemma~\ref{eta''}. 
\hfill$\Box$
\subsection{Proof of theorem~\ref{global}}
We can continue to obtain bounds on higher derivatives of the fields 
by repeating the above arguments. 
Fortunately, to apply the global existence theorem in \cite{MA}, 
it is enough to get $C^2$ bounds of all 
functions. Thus, it has been shown that 
the functions $\eta$, $\alpha$, $U$, $A$, $\phi$ and $\sigma$ extend to 
$t\rightarrow 0$ into the past direction and to $t\rightarrow \infty$ into 
the future direction. 
\hfill$\Box$
\section{Comments}
We should like to comment concerning the TCC and 
the condition for coupling constant $\lambda$. 
Note that these conditions are needed to prove theorem~\ref{global} into 
only the past direction. 
It is expected that the TCC is satisfied near initial singularities 
because strong focusing effect by gravity is dominant than repulsing one by 
a positive potential (cosmological constant) there. 
Note that spacetimes described by our AVTD solutions satisfy the TCC. 
Also, it is possible to expand in acceleration of the spacetimes 
into the future direction since the TCC does not hold necessarily there 
and the positive potential would become dominant. 
Thus, theorem~\ref{global} does not deny paradigm-A. 

The condition $\left|\lambda\right|\leq\bar{\lambda}<1/2$ admits $\lambda=0$, 
which means that there is a positive cosmological constant. 
Thus, our theorem is applicable to not only theories with dilaton coupling 
but also ones with a pure cosmological constant. 
Now, let us discuss $\bar{\lambda}$. 
It is known that there is a critical value $\lambda_{C}$
in $n$-dimensional homogeneous and isotropic spacetimes~\cite{TP,WMNR}. 
In our notation with $n=4$, $\left|\lambda_{C}\right|=1/2$.
Here, "critical" means the boundary whether late-time attractor solutions 
indicate accelerated expansion or not. 
Roughly speaking, $\lambda$ describes steepness of the potential. 
Therefore, for $\lambda^2>\lambda_{C}^2$, the dilaton field falls down the 
potential hill soon and then decelerating expansion solutions with transient 
accelerating one are obtained, while we have attractor solutions with eternal 
accelerating expansion if $\lambda^2<\lambda_{C}^2$. 
It is believed that such critical values exist for generic spacetimes, 
although we do not know $\lambda_{C}$ for spacetimes we considered here, 
in particular, our results give us no information about 
relation between $\lambda_{C}$ and $\bar{\lambda}$. 
Thus, it is not clear that the solution obtained 
in theorem~\ref{avtd2} is consistent with paradigm-A at 
the intermediate- and late-time. 
To answer this question, we need 
to analyze future asymptotic behavior (e.g. see~\cite{RA04}), which 
is left for future research.

\begin{center}
{\bf Acknowledgments}
\end{center}
I am grateful to Alan Rendall
and Yoshio Tsutsumi 
for commenting on the manuscript.

\appendix
\section{Local existence and uniqueness for smooth case}
Let us consider the smooth version of the initial-value problem 
for our non-standard setup formulated in section \ref{initial-singularity}. 
A key idea is to construct a symmetric-hyperbolic system by introducing 
a new variable $\alpha':=Z_{14}$~\cite{IK}. 
Let us define $\vec{Z}:=Z_{i}=(U,\dot{U},U',A,\dot{A},A',\phi,\dot{\phi},\phi',
\sigma,\dot{\sigma},\sigma',\alpha,\alpha',\eta)$. 
Here, $i$ runs from $1$ to $15$. 
The system consisting in the effective evolution equations 
(\ref{ce-eta}), 
(\ref{ce-alpha}), 
(\ref{ee-u})-(\ref{ee-sigma}) becomes the following first-order 
symmetric-hyperbolic one: 
\begin{eqnarray}
{\cal A}_{0}\partial_{t}\vec{Z}={\cal A}_{1}\partial_{\theta}\vec{Z}+
F(t,\theta,\vec{Z}), 
\end{eqnarray} 
where 
\begin{eqnarray}
{\cal A}_{0}={\rm diag}(1,1,\alpha,1,1,\alpha,1,1,\alpha,1,1,\alpha,1,1,1),
\end{eqnarray} 
and 
\begin{eqnarray}
{\cal A}_{1}=
\left(\begin{array}{ccccc}
{\cal A}_{2} & 0 & 0 & 0 & 0\\
0 & {\cal A}_{2} & 0 & 0 & 0\\
0 & 0 & {\cal A}_{2} & 0 & 0\\
0 & 0 & 0 & {\cal A}_{2} & 0\\
0 & 0 & 0 & 0 & {\cal A}_{3}\\
\end{array}\right),
\end{eqnarray} 
\begin{eqnarray}
{\cal A}_{2}=
\left(\begin{array}{ccc}
0 & 0 & 0\\
0 & 0 & \alpha\\
0 & \alpha & 0\\
\end{array}\right) 
\hspace{.5cm}{\rm and}\hspace{.5cm}
{\cal A}_{3}=
\left(\begin{array}{ccc}
0 & 0 & 0\\
0 & 0 & 0\\
0 & 0 & 0\\
\end{array}\right). 
\end{eqnarray} 
Thus, we have a unique 
solution to the effective evolution equations by prescribing the smooth 
initial data for $t=t_{0}>0$ if the constraint equations (\ref{ce-eta'}) 
and $\alpha'=Z_{14}$ hold for any $t$.  

Now, as the analytic case,  
to assure the local existence and uniqueness of the initial-value problem, 
it is enough to show that the constraints propagate.  
Let us set 
\begin{eqnarray}
N_{1}:=\eta'-2DUU'-\frac{e^{4U}}{2t^2}DAA'-\frac{1}{2}D\phi\phi'
-\frac{e^{2\lambda\phi}}{2}D\sigma\sigma'+\frac{Z_{14}}{2\alpha},
\end{eqnarray} 
and 
\begin{eqnarray}
N_{2}:=Z_{14}-\alpha'.
\end{eqnarray} 
By direct calculation, we have the following linear, homogeneous ODE system: 
\begin{eqnarray}
(D+{\cal B})\vec{N}=0, 
\end{eqnarray} 
where $\vec{N}:=(N_{1},N_{2})$ and 
\begin{eqnarray}
{\cal B}=\frac{D\alpha}{2\alpha^2}
\left(\begin{array}{cc}
\alpha & -1 \\
4\alpha^2 & -2\alpha\\
\end{array}\right).
\end{eqnarray} 
Thus, the uniqueness theorem for ODE systems guarantees that $\vec{N}$ 
is identically zero 
for any time $t$ if we set initial data for $t=t_{0}$ such that 
$\vec{N}(t_{0})=0$. 
Thus, the local existence and uniqueness 
of the initial-value problem for our case 
has been shown in the smooth case.


\end{document}